\documentclass[11pt]{article}
\usepackage{geometry}                
\usepackage{graphicx}
\usepackage{amssymb}
\usepackage{epstopdf}
\newcommand{\be}{\begin{eqnarray}}
\newcommand{\ee}{\end{eqnarray}}
\title{Shared information between residues is sufficient to detect pair-wise epistasis in a protein}
\author{Aditi Gupta$^1$ \& Christoph Adami$^{2,3,4,\star}$}

\begin{document}
\maketitle

\begin{center}
$^1$Center for Infectious Diseases, New Jersey Medical School\\
 Rutgers University, Newark, New Jersey, USA\\
$^{2}$Department of Microbiology \& Molecular Genetics\\
$^{3}$Department of Physics and Astronomy\\
$^{4}$BEACON Center for the Study of Evolution in Action\\
Michigan State University, East Lansing, MI 48824\\
\end{center}
$^\star$Corresponding author: adami@msu.edu
\vskip 2cm
In a comment~\cite{Crona2016} on our manuscript ``Strong selection significantly increases epistatic interactions in the long-term evolution of a protein"~\cite{GuptaAdami2016}, Dr. Crona challenges our assertion that shared entropy (that is, information) between two residues implies epistasis between those residues, by constructing an explicit example of three loci (say A, B, and C), where A and B are epistatically linked (leading to shared entropy between A and B), and A and C also depend epistatically (leading to shared entropy between A and C), so that loci B and C are correlated (share entropy). She goes on to assert that (as per her examples), even though there will be correlations (and thus shared entropy) between the residues at loci B and C, there is no pair-wise epistasis between loci B and C, contradicting our assertion in~\cite{GuptaAdami2016} that shared entropy implies epistasis. 

The disagreement is based on two different interpretations of the meaning of pair-wise epistasis, and the comment gives us an opportunity to discuss those.

We do not disagree that epistasis refers to mutational effects that are conditional on the states of other alleles. In our paper, we are in particular interested in pair-wise epistasis, that is, how the fitness effects of mutations at two loci depend on each other. Of course, the dependence between those two loci could depend on the state of many other alleles in the genome. The different interpretations of epistasis hinge upon whether a quantitative assessment of the epistasis between two loci should be conditional on the state of other loci in the genome, or whether instead we should consider the state of these loci averaged over what their state would be in a population at mutation-selection balance. Crona's example helps us illustrate that distinction. 

A fitness landscape constructed in the supplementary information of~\cite{Crona2016} (Example 1) has three loci, A, B, and C. The values given are (note that in~\cite{Crona2016}, the log-transformed relative fitnesses are given instead):
\be
w_{000}=1, w_{001}=0.1, w_{010}=0.1, w_{011}=0.01, w_{100}=1.1, w_{101}=1.1, w_{110}=1.21, w_{111}=1.331\;.
\ee
The values are judiciously chosen such that 
\be
\frac{w_{000}\,w_{011}}{w_{010}\,w_{001}}=\frac{w_{100}\,w_{111}}{w_{110}\,w_{101}}=1\;.
\ee 
Crona defines the epistasis between loci B and C as either
\be
\epsilon_{BC}^{A=0}=\log\left(\frac{w_{000}\,w_{011}}{w_{010}\,w_{001}}\right)
\ee
or 
\be
\epsilon_{BC}^{A=1}=\log\left(\frac{w_{100}\,w_{111}}{w_{110}\,w_{101}}\right)\;.
\ee
Note that these values are conditional on the state of locus A, but for the case discussed here they happen to be equal, and vanishing. Clearly, this is a special case. In general, the pair-wise epistasis conditional on the state of another locus can depend on that state, and if there are $n$ other binary loci, then there could be in principle $2^n$ different values for the pair-wise epistasis. Surely this is not satisfactory, as pair-wise epistasis then would not be defined. Instead, pair-wise epistasis should be {\em unconditional} on the state of other loci in the genome. How do you calculate this?

We assert that pair-wise epistasis between two loci should depend on the fitness effect of mutations at those loci, but where the states of the other loci are determined by mutation-selection balance in a population. In other words, we assert that fitness effects should be measured by the effect on the growth rate of a population. For the three locus system, the fitness of the BC system depends on the frequency of the A=0 allele and the A=1 allele in the population. Let $p_0$ stand for the frequency of the A=0 allele, with $p_0+p_1=1$. Then
\be
w_{00}&=&p_0w_{000}+p_1 w_{100}\;,\\
w_{01}&=&p_0w_{001}+p_1 w_{101}\;,\\
w_{10}&=&p_0w_{010}+p_1 w_{110}\;,\\
w_{11}&=&p_0w_{011}+p_1 w_{111}\;.
\ee                                       
These four values can be used to calculate the epistasis between loci B and C unconditional on the state of A as
\be
\epsilon_{BC}=\log\left(\frac{w_{00}\,w_{11}}{w_{10}\,w_{01}}\right)\;.
\ee
We plot this quantity in Fig. 1 as a function of the frequency of the A=0 allele $p_0$, and see that it is everywhere positive except for  $p_0=0$ or $p_0=1$, which are the conditional epistasis values of Crona, Eqs. (2) and (3). 
\begin{figure}[htbp] 
   \centering
   \includegraphics[width=4in]{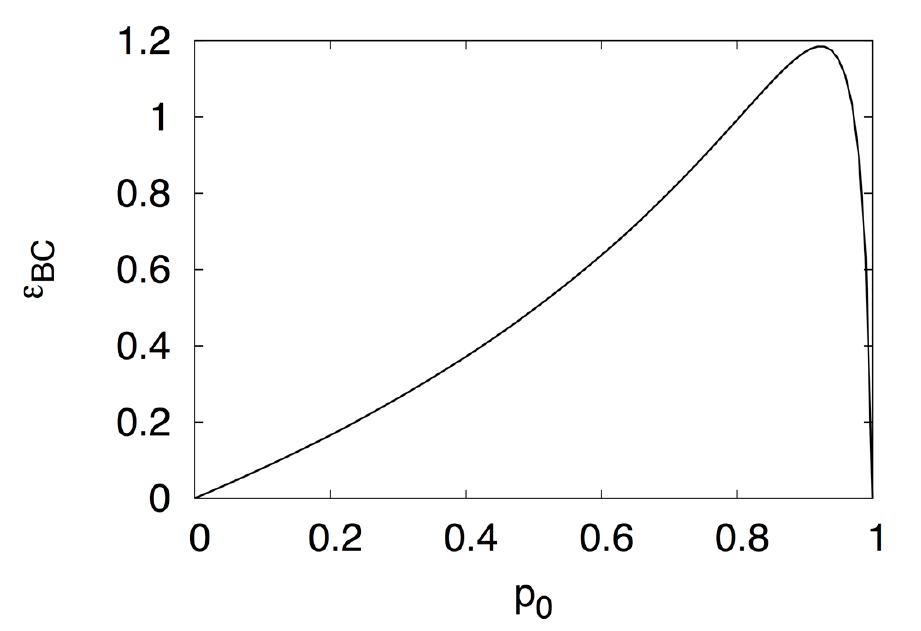} 
   \caption{{\bf Unconditional epistasis.} Pair-wise epistasis between loci B and C calculated using Eqs. (4-8) and (9), as a function of locus A allele frequency $p_0$, with fitnesses as in Eq. (1).}
   \label{fig:example}
\end{figure}
However, given the fitness landscape Eq. (1), these extreme values (a population composed purely of one allelic state of A) are impossible. As long as the mutation rate is non-vanishing there will always be a mixture of both alleles at locus A. Indeed, Table 1 of the supplementary information of~\cite{Crona2016}, which tabulates an evolutionary simulation on that precise landscape, makes that point for us. Crona finds that  $p_0\approx 0.998$  in equilibrium, leading to $\epsilon_{BC}\approx0.191$, which is non-vanishing. Thus, the positive shared entropy between those loci is indeed sufficient to determine non-vanishing pair-wise epistasis between them. We also remark that at that frequency $p_0$, the information between B and C is exceedingly small: $I\approx0.0013$.

Needless to say, the example discussed here is a fairly contrived one, and we show in Fig.\ 9 of~\cite{GuptaAdami2016} that the correlation between epistasis and information is robust when testing random fitness landscapes. The same argument holds for examples 2 and 3 in the comment~\cite{Crona2016}.  

In conclusion, the assertion in~\cite{Crona2016} that detecting epistasis via shared entropy gives false positives for epistasis is based on a calculation of conditional epistasis, a concept that is ambiguous at best as it depends on the allelic state of all the other loci on the genome, and could take on arbitrary values.  If epistasis is calculated by averaging over the allelic state of the other loci, then our assertion that correlation (positive information) implies positive pair-wise epistasis holds without exception. 
\vskip 0.5cm
\noindent{\bf Acknowledgements}
This research has been supported in part by the National Science Foundation (NSF) BEACON Center under Cooperative Agreement DBI-0939454.

\end{document}